\documentclass[12pt]{article}

\usepackage{amsmath,amssymb}
\usepackage{xspace}
\usepackage{graphicx}
\usepackage{caption}
\usepackage{color}
\usepackage{dcolumn}
\usepackage{bm}

\usepackage[numbers,super,comma,sort&compress]{natbib}

\definecolor{background-color}{gray}{0.98}

\graphicspath{{figs/}}

\def\bA{{\bf A}\xspace}
\def\bB{{\bf B}\xspace}

\def\bE{{\bf E}\xspace}

\def\bH{{\bf H}\xspace}
\def\bS{{\bf S}\xspace}

\def\bSig{\boldsymbol{\Sigma}\xspace}
\def\bDel{\boldsymbol{\Delta}\xspace}

\def\bP{{\bf P}\xspace}
\def\bQ{{\bf Q}\xspace}

\def\bU{{\bf U}\xspace}
\def\bV{{\bf V}\xspace}
\def\bX{{\bf X}\xspace}
\def\bY{{\bf Y}\xspace}
\def\bZ{{\bf Z}\xspace}

\def\bb{{\bf b}\xspace}

\def\bsig{\boldsymbol{\sigma}\xspace}
\def\btau{\boldsymbol{\tau}\xspace}

\def\bx{{\bf x}\xspace}
\def\bu{{\bf u}\xspace}

\def\br{{\bf r}\xspace}
\def\ham{\hat{H}}

\def\w{\omega\xspace}
\def\bra#1{\langle #1|}
\def\ket#1{|#1\rangle}
\def\braket<#1|#2>{\langle #1| #2 \rangle}
\def\TDA{\text{TDA}}
\def\RPA{\text{RPA}}

\title{
 Simulating X-ray absorption spectra with CASSCF linear response methods
}
\author{
 Benjamin Helmich-Paris \thanks{Max-Planck-Institut f{\"u}r Kohlenforschung, D-45470 M{\"u}lheim an der Ruhr}
}

\begin{document}

\maketitle

\begin{abstract}
In this work, two approaches for simulating
X-ray absorption (XA) spectra with the complete active space self-consistent field
(CASSCF) linear response (LR) method are introduced.
The first approach employs the well-known core-valence separation (CVS) approximation,
which is predominantly used by many other electronic structure methods for simulating
X-ray spectra.
The second ansatz uses the harmonic Davidson algorithm (DA) for finding interior eigenvalues
that lie close to a target excitation energy shift and virtually solves
a shifted-and-inverted (S\&I) generalized eigenvalue problem.
Our implementations of these Davidson-type algorithms for core spectroscopies
converge as rapidly as the standard DA for valence excitations.
It is shown in a proof-of-principle application to the Manganese atom that the
additional errors for excitations energies introduced by the CVS approximation compared to the exact S\&I approach
are negligible for K-edges, but become larger than the methodological error
of the CASSCF LR method when computing M-edges.
We could also demonstrate that, in case of large active-space K-edge calculations, the computational savings of
the CVS approximation are huge as the configuration part of the response vectors
is completely neglected.
When simulating the oxygen K-edge XA spectrum of the permanganate ion,
CASSCF LR showed a better agreement with the experimental spectrum than the CAS-CI and NEVPT2 methods,
 however, the order of the two close-together pre-edge peaks seems to be inverted.
\end{abstract}

\clearpage


  \makeatletter
  \renewcommand\@biblabel[1]{#1.}
  \makeatother

\bibliographystyle{apsrev}

\renewcommand{\baselinestretch}{1.5}
\normalsize

\clearpage

\section{Introduction}

In the last decades, the field of theoretical and computational spectroscopy has made enormous progress
and allows us today to simulate various types of experimental X-ray spectra.\cite{Norman2018}
Since the initial work of Bagus in 1965 on core-hole states with Hartree--Fock theory,\cite{Bagus1965}
many milestone have been reached in the course of time as 
the core-valence separation (CVS) approximation,\cite{Cederbaum1980}
the maximum overlap method,\cite{Gilbert2008}
real-time propagation approaches,\cite{Akama2010,Lopata2012,Kadek2015,Repisky2015}
complex polarization propagators,\cite{Norman2005}
oscillator strengths beyond the electric dipole approximation,\cite{DeBeer2008,Bernadotte2012,List2015,Soerensen2019}
etc.

In particular for transition-metal (TM) compounds X-ray spectroscopy is a frequently used 
technique that
can provide valuable insights into the electronic structure, oxidation and spin state, and
coordination number of the metal centers.\cite{Sinfelt1993}
Those open-shell TM compounds show a characteristic pre-edge peak in the X-ray absorption (XA) spectrum
stemming from the excitation of an electron from a core shell into the partially filled TM 3d shell.
However, open-shell TM complexes usually have a complicated electronic structure
that can only be described properly either by a restricted open-shell ansatz\cite{Roemelt2013,Roemelt2013a} or by
multi-reference (MR) methods.\cite{Josefsson2012,Kunnus2013,Pinjari2014}
The latter are unavoidable when calculating open-shell TM complexes in a low-spin state.

Similar multi-step protocols based on MR methods
have been developed by several groups to simulate XA spectra of TM complexes.\cite{Josefsson2012,Pinjari2014,Chantzis2018,Manganas2019}
These approaches can be briefly classified into the following four steps:
(i) Orbitals for the electronic ground state are obtained from a 
    complete active space self consistent field (CASSCF) calculation
    either for a single state or for an average of a few low-lying states.
    Usually, the valence TM d and optionally the attached ligand electrons and orbitals are included in the active space.
(ii) Then, a complete or restricted active space (RAS) configuration interaction (CI) calculation 
    is performed for the core-excited states with an active space that 
    has been extended by the desired core electrons and orbitals.
    In this way, the CVS approximation is easily implemented
    that restricts the space of CI determinants to those
    describing excitations from the core into the valence orbital subspace.
(iii) In case of L- and M-edge XA spectra, spin-orbit coupling (SOC) must be
    included for describing the splitting of the 2p and 3p shells, respectively, and
    to obtain even qualitatively correct results.
    Those multi-step approaches usually employ state interactions\cite{Malmqvist2002,Roemelt2013a} (SI) of the
    nonrelativistic core-excited RAS-CI states with different multiplicities that are 
    coupled by an (effective) one-electron SOC perturbation operator\cite{Hess1996,Schimmelpfennig1998}.
(iv) Dynamic correlation is often introduced by performing MR second-order perturbation theory\cite{Andersson1990,Andersson1992,Finley1998,Angeli2001,Angeli2001b,Angeli2002,Angeli2004} (PT2)
    or MR coupled cluster\cite{Datta2012,Brabec2012,Dutta2014a} (CC)
    on top of the many nonrelativistic core-excited RAS-CI wavefunctions.
    When simulating L- or M-edge spectra, the state energies of such accurate calculations 
    are then incorporated into the SOC SI Hamiltonian by means of diagonal shifts.\cite{Malmqvist2002}

In recent years, the applicability and accuracy of such type of multi-step approaches
has been successfully demonstrated by several computational studies.\cite{Josefsson2012,Kunnus2013,Pinjari2014,Chantzis2018,Manganas2019}
Nevertheless, it should be mentioned that such multi-step approaches are usually
only suited to compute the core-excitation energy fairly accurately.
Compromises are made already when computing oscillator strengths or the SOC SI
Hamiltonian for which transition or SOC matrix elements are computed at the CASSCF / RAS-CI level.
One has to face similar limitation when computing other properties related
to core excitations following such multi-step protocols.

The current work focuses on a one-step approach for simulating XA spectra with
MR electronic structure methods.
Our approach is based on computing core excitation energies and oscillator
strengths from the CASSCF linear response (LR) function\cite{Yeager1979,Dalgaard1980,Olsen1985,Joergensen1988,Helmich-Paris2019}
and builds upon previous work on CASSCF LR valence electron absorption spectra for large molecules.\cite{Helmich-Paris2019}
With the LR or polarization propagator (PP) ansatz, excitation energies
and oscillator strengths are determined from the poles and residues of the
LR function,\cite{Olsen1985,Christiansen1998}
which can be used for simulating valence and core excitations alike.
The preference of LR theory over approaches that
compute excited states directly by a state averaging and/or CI procedure is justified by
(i) higher accuracy due to state-specific eigenvectors representing excited state wavefunctions,
(ii) the avoidance of artificial valence-Rydberg mixing with diffuse basis sets,
(iii) and the capability of describing excited states that are not exclusively represented
     by the active space CI determinants.
The latter point becomes relevant when simulating the the near-edge fine structure 
of the XA spectrum that, in the one-particle picture, corresponds to electronic excitations from the core
into the low-lying unoccupied orbitals.
We note at this point that the near K-edge fine structure features of the XA spectra 
of closed-shell atoms and small molecules were calculated with impressive accuracy
using CC LR theory.\cite{Coriani2012}
Most likely, one can expect a similar accuracy for open-shell systems
once highly accurate MR-PT and MR-CC LR methods are available for core excitations.

For single-reference (SR) methods a combination of LR and PP
approaches with the 
CVS approximation is well established and goes back
to the early work of Barth and Schirmer\cite{Barth1985} on K-edge spectra using 
the second-order algebraic diagrammatic construction\cite{Schirmer1982}
in combination with the CVS approximation.\cite{Cederbaum1980}
Since then, the CVS approximation has been employed along with many SR LR
and related methods as CC LR\cite{Coriani2012,Coriani2015} and
equation-of-method (EOM) methods,\cite{Vidal2019}
LR time-dependent density functional theory (DFT),\cite{DeBeer2008a,DeBeer2008,Besley2010,Lopata2012}
or the doubles (D) correction to configuration interaction singles (CIS).\cite{Asmuruf2008}

Another strategy for determining core excitations with LR approaches
is to use modified iterative diagonalization algorithms designed for
finding interior eigenvalues of LR eigenvalue equations.
In the present work, the harmonic Davidson method of Morgan\cite{Morgan1991,Morgan1998}
is employed that searches for all interior eigenvalues close to a target 
excitation energy shift.
Compared to even more sophisticated iterative diagonalization algorithms,\cite{Fokkema1998,Vecharynski2015,Zuev2015,Kasper2018}
only modest adaptions of the original multi-state Davidson algorithm\cite{Davidson1975} (DA) for the
lowest excitation energies become necessary.
The harmonic DA has been employed already in several 
other fields of electronic structure theory, e.g.\ for finding a few selected
valence bands in solids using Kohn-Sham DFT.\cite{Tackett2002,Jordan2012}
Also for density matrix renormalization group methods
for quantum chemistry\cite{Dorando2007,Baiardi2019}, 
the harmonic DA is advantageous because the more accurate state-specific 
ansatz or averaging over fewer states becomes available.

This article shows how the CASSCF LR approach,
also known as the multiconfigurational random phase approximation (MC-RPA), 
can be employed for simulating XA spectra
using either the CVS approximation or the harmonic DA.
General aspects of CASSCF LR theory with a focus on computing
excitation energies are presented in Sec.\ \ref{sec:theory}.
Moreover, the CVS approximation for MC-RPA calculations and
the MC Tamm--Dancoff approximation (MC-TDA) is presented in more
detail and key aspects of the harmonic DA
for MC-TDA and MC-RPA calculations are covered.
After providing technical details on the calculations in Sec.\ \ref{sec:compdet}, results on proof-of-principle
calculations are shown in Sec.\ \ref{sec:results}.
Here, we mainly focus on accuracy of the CVS approximation, convergence of the (harmonic) Davidson
algorithm, and accuracy of the MC-RPA method 
in comparison with other MR methods when be applied for K-edge spectra.

\section{Theory}\label{sec:theory}

\subsection{CASSCF linear response approach} \label{sec:DA}

The CASSCF method is a fully variational wavefunction model. 
For those models, the time-dependent (TD) phase-isolated wavefunction
can be obtained from the time-independent wavefunction $\ket{0}$
by unitary transformation.\cite{Dalgaard1980,Olsen1985,Christiansen1998}
Propagating the CASSCF wavefunctions in time is established by unitary
transformations
 of the orbital and configuration part of $\ket{0}$
\begin{align}\label{eq:td-wf}
  \ket{\tilde{0}} &= \exp( \hat{\kappa}(t)) \, \exp( \hat{S}(t)) \, \ket{0}
\text{.}
\end{align}
The exponential form of the TD unitary transformations in Eq.\ \ref{eq:td-wf}
demands the exponents to be anti-Hermitian TD operators.
The exponent of the orbital part is given by TD orbital rotations that involve
single-electron excitation $\hat{q}^{\dag}_m$ and deexcitation operators $\hat{q}^{\dag}$,
\begin{align} \label{eq:td-wf-oc}
 &\hat{\kappa}(t) = \sum_m     \left( \kappa_m(t) \, \hat{q}^{\dag}_m - \kappa_m^*(t) \, \hat{q}_m \right) \text{,} \\
 &\begin{array}{ccc}
 \hat{q}^{\dag}_m = \hat{E}_{pq} = \hat{a}^{\dag}_{p \alpha} \hat{a}_{q \alpha} + \hat{a}^{\dag}_{p \beta} \hat{a}_{q \beta} &
\text{ and }
&
 \hat{q}_m = \hat{E}_{qp}
 \end{array}
\text{,}
\end{align}
while the exponent of the configuration part involves 
state-transfer operators $\hat{R}_m^{\dag}$ and $\hat{R}_m$ between 
the time-independent reference state CI wavefunction $\ket{0}$ and
the orthogonal complement states $\ket{m}$ 
\begin{align}
 &\hat{S}(t) = \sum_{m>0} \left( S_m(t)      \, \hat{R}^{\dag}_m - S_m^*(t)      \, \hat{R}_m \right) \text{,} \\
 &\begin{array}{ccc}
 \hat{R}_m^{\dag} = \ket{m}\bra{0}
&
\text{ and }
& \hat{R}_m        = \ket{0}\bra{m} 
 \end{array}
\text{,}
\end{align}
that, if combined with $\ket{0}$,
form the complete set of CAS-CI configuration state functions (CSF).

In response theory, the external electromagnetic field is
treated as time-periodic perturbation.\cite{Olsen1985,Christiansen1998,Helgaker2012}
Applying the time-dependent variational principle 
leads in second order in the perturbation strengths to the generalized eigenvalue equations
\begin{align} \label{eq:eigrsp}
    & \bE^{(2)}\, \bX = \w \, \bS^{(2)}\, \bX
\end{align}
that involve the electronic Hessian $\bE^{(2)}$ and a generalized metric $\bS^{(2)}$ matrices.
The solution of Eq.\ \ref{eq:eigrsp} gives the poles of the CASSCF linear
response function $\w$, which can be identified as excitation energies in an exact theory.\cite{Olsen1985,Christiansen1998}
The duality of excitation and deexcitation operators to create anti-Hermitian
operators $\hat{\kappa}(t)$ and $\hat{S}(t)$ leads to a 
two-by-two block structure of the $\bE^{(2)}$ and $\bS^{(2)}$ matrices,
 \begin{align}
  \begin{pmatrix}
   \bA   & \bB \\
   \bB^* & {\bA}^*
  \end{pmatrix}
  \begin{pmatrix}
    \bZ \\
    \bY^*
  \end{pmatrix}
=
  \w
  \begin{pmatrix}
    \bSig   &  \bDel  \\
   -\bDel^* & -\bSig^*
  \end{pmatrix}
    \begin{pmatrix}
      \bZ \\
      \bY^*
    \end{pmatrix}
\text{.}
    \label{eq:eigparti}
\end{align}
Due to block structure of Eq.\ \ref{eq:eigparti}, there are two different solutions
of the eigenvalue problem
\begin{align}
&
 \begin{array}{ccc}
 +\w  & \text{ and } & \bX(+\w) =
 \begin{pmatrix}
  \bZ \\
  \bY^*
 \end{pmatrix} \text{,} \\
-\w & \text{ and } & \bX(-\w) = 
\begin{pmatrix}
  \bY \\
  \bZ^*
 \end{pmatrix}  \text{,}
 \end{array}
\end{align}
representing electronic excitations and de-excitations, respectively. 
In order to not break the paired structure of these RPA-type eigenvalue problems
both solutions need to be determined simultaneously.
Alternatively, linear (plus and minus) combinations of the 
eigenvectors $\bX(\w)$ and $\bX(-\w)$,
\begin{align} \label{eq:vec-combi}
   {\bU}(h) &= \bX(+\w) + h \,  \bX(-\w) =
 \begin{pmatrix}
  \bZ   + h \, \bY \\
  \bY^* + h \, \bZ^*
 \end{pmatrix}
  = 
 \begin{pmatrix}
  \bV \\
 h \, \bV^*
 \end{pmatrix}
\text{,}
 \end{align}
can be determined rather than the eigenvectors themselves,
which is known as the Hermitian ($+h$) and anti-Hermitian ($-h$)
form of the eigenvectors introduced by Saue and Jensen.\cite{Saue2003}

In practice, we are only interested in a few roots of the LR eigenvalue
problem, which are determined iteratively by Davidson-type algorithms.\cite{Davidson1975}
For such iterative diagonalization algorithms,
products of a few trial vectors $\bb$ with the large matrices $\bE^{(2)}$ and $\bS^{(2)}$
of Eq.\ \ref{eq:eigrsp} need to be computed. 
As was shown originally by Saue and Jensen,\cite{Saue2003,Helmich-Paris2019} 
the Hermicity of the Hermitian ($h = +1$) 
and anti-Hermitian ($h = -1$) trial vectors is conserved when being
multiplied with the Hermitian $\bE^{(2)}$ matrix,
  \begin{align} \label{eq:e-trafo}
    \begin{pmatrix}
     \bA   & \bB \\
     \bB^* & {\bA}^*
    \end{pmatrix}
    \begin{pmatrix}
      \bb \\
 h \, \bb^*
    \end{pmatrix}
    =
    \begin{pmatrix}
    \bA \bb + h\, \bB \bb^* \\
    h\, (\bA \bb + h\, \bB \bb^*)
     \end{pmatrix}
    = \bsig_{( h )}
\end{align}
and inverted when being multiplied with the anti-Hermitian $\bS^{(2)}$ matrix
    \begin{align} \label{eq:s-trafo}
    &
    \begin{pmatrix}
      \bSig   &  \bDel  \\
     -\bDel^* & -\bSig^*
    \end{pmatrix}
        \begin{pmatrix}
      \bb \\
 h \, \bb^*
    \end{pmatrix}
    = 
  \begin{pmatrix}  
   \bSig \bb + h\, \bDel \bb^* \\
   -h\, ( \bSig \bb + h\, \bDel \bb^* )
  \end{pmatrix}
  = \btau_{( -h )}
 \text{,}
    \end{align}
As becomes apparent from Eqs.\ \ref{eq:e-trafo} and \ref{eq:s-trafo}, only one component (upper or lower) 
of the (anti-)Hermitian trial vectors $\bb$ and linearly transformed 
trial vectors $\bsig$ and $\btau$ is sufficient. 
We choose the upper component without loss of generality. 
If needed, the redundant lower component is readily available by complex conjugation and an
additional  phase change for anti-Hermitian vectors.

One of the advantages of exploiting Hermicity emerges when computing vector products.
If two vectors with them same Hermicity ($h1 = h2$) are multiplied, their product is purely real;
if two vectors with different Hermicity ($h1 \ne h2 $) are multiplied, their product is purely imaginary,
\begin{align}
 \bU^{\dag}_{1,(h_1)}\, \bU_{2,(h_2)} &= \bV_1^{\dag}\,\bV_2 + h_1 h_2 (\bV_1^{\dag}\, \bV_2)^* \text{,} \\
   &= \left\{
   \begin{matrix}
      2  \, \text{Re} (\bV_1^{T}\,\bV_2 ) \text{ if } h_1 =   h_2 \\
      2i \, \text{Im} (\bV_1^{T}\,\bV_2 ) \text{ if } h_1 \ne h_2  
   \end{matrix} \right.
 \text{.}
\end{align}
This becomes relevant when formulating the MC-RPA eigenvalue problem, Eq.\ \ref{eq:eigparti}, in the reduced space spanned by the trial vectors $\bb$.
Since we are dealing only with real quantities when solving the LR eigenvalue
equations for non-relativistic Hamiltonians, two out of the four blocks of the reduced space $\bE^{[2]}$ and $\bS^{[2]}$
matrices vanish
\begin{align} \label{eq:mcrpa-eig-red}
\begin{pmatrix}
  \bb_{(+)}^{\dag} \bsig_{(+)}  & 0\\
  0 & \bb_{(-)}^{\dag} \bsig_{(-)}
\end{pmatrix}
\begin{pmatrix}
\bu_{(+)} \\
\bu_{(-)}
\end{pmatrix}
=
\w_i
\begin{pmatrix}
 0 &  \bb_{(+)}^{\dag} \btau_{(+)} \\
 \bb_{(-)}^{\dag} \btau_{(-)} & 0
\end{pmatrix}
\begin{pmatrix}
\bu_{(+)} \\
\bu_{(-)}
\end{pmatrix}
\text{.}
\end{align}
In a previous work on UV/Vis absorption spectra with the MC-RPA method,\cite{Helmich-Paris2019} 
this Hermicity-block structure was exploited and lead to substantial computational savings
if many roots are optimized simultaneously.

\subsection{Core-valence separation approximation} \label{sec:cvs}

The sub-blocks of $\bE^{(2)}$ and $\bS^{(2)}$ appearing in Eq.\ \ref{eq:eigparti} can be decomposed
further according to their operator types in the wavefunction model.
In case of CASSCF-type wavefunction models,
there is in turn a two-by-two block structure when separating the 
orbital $\hat{\kappa}$ and configuration $\hat{S}$ rotation part
in the super matrices $\bA$, $\bB$, $\bSig$, and $\bDel$,
\begin{align}
 \begin{array}{ll}
  A_{ij} = &
   \begin{pmatrix}
    \bra{0} [ \hat{q}_i , [\ham, \hat{q}_j^{\dag} ]] \ket{0} & 
    \bra{0} [ \hat{q}_i , [\ham, \hat{R}_j^{\dag} ]] \ket{0} \\
    \bra{0} [ \hat{R}_i , [\ham, \hat{q}_j^{\dag} ]] \ket{0} & 
    \bra{0} [ \hat{R}_i , [\ham, \hat{R}_j^{\dag} ]] \ket{0}
   \end{pmatrix} 
\text{,}
\\[1.5em]
 B_{ij}= &
  \begin{pmatrix}
   \bra{0} [  q_i , [\ham, q_j ]] \ket{0} & 
   \bra{0} [  q_i , [\ham, R_j ]] \ket{0} \\
   \bra{0} [  R_i , [\ham, q_j ]] \ket{0} & 
   \bra{0} [  R_i , [\ham, R_j ]] \ket{0}
  \end{pmatrix} 
\text{,}
\\[1.5em]
 \Sigma_{ij} = &
  \begin{pmatrix}
   \bra{0} [ q_i, q_j^{\dag} ] \ket{0} & 
   \bra{0} [ q_i, R_j^{\dag} ] \ket{0} \\
   \bra{0} [ R_i, q_j^{\dag} ] \ket{0} & 
   \bra{0} [ R_i, R_j^{\dag} ] \ket{0}
  \end{pmatrix}
\text{,}
\\[1.5em]
 \Delta_{ij} =&
  \begin{pmatrix}
   \bra{0} [ q_i, q_j ] \ket{0} & 
   \bra{0} [ q_i, R_j ] \ket{0} \\
   \bra{0} [ R_i, q_j ] \ket{0} & 
   \bra{0} [ R_i, R_j ] \ket{0}
  \end{pmatrix}
\text{.}
\end{array}
\end{align}
The structure of the Hessian matrices $\bA$ (or $\bB$)
is shown schematically in Fig.\ \ref{fig1}.
The orbital-orbital block can be decomposed further into
the three possible types of single-particle excitations between the
inactive occupied (O), active (A), and virtual (V) MO spaces.
All Hessian matrix blocks have nonvanishing elements.

Excitation energies associated with an excitation from the
core into the partially occupied active or unoccupied virtual orbitals
occur as interior eigenvalues of the Hessian spectrum.
For molecular systems with a decent basis set size 
it is computationally intractable to find those eigenvalues
by a full or iterative diagonalization of the 
full Hessian (and generalized metric) matrix.

A well-established strategy for obtaining XA spectra with LR or PP methods is 
to restrict the search space to only those vector elements
that have a few application-relevant core orbitals.
In most cases, coupling of the core and valence part is usually weak
and is, thus, neglected within this so-called CVS approximation.

The CVS CASSCF Hessian (and generalized metric) matrix only has
non-vanishing elements in the  virtual-occupied (V--O) and active-occupied (A--O) orbital part. 
Only elements with one of the selected core orbitals enter the V--O and A--O blocks.
The configuration part is neglected completely because we
assume that for one-step MR LR or PP approaches the active space usually comprises valence orbitals only.
The block structure of the CVS CASSCF Hessian is illustrated in Fig.\ \ref{fig2}.
The energetically lowest eigenvalues of the CVS CASSCF Hessian and metric
correspond altogether to core excitations and can be easily found by
the same DA used for valence excited states (see Sec.\ \ref{sec:DA}).
Note that the computational costs for diagonalizing the CVS CASSCF Hessian and
generalized metric can be significantly smaller in cases of larger
active spaces as will be demonstrated in Sec.\ \ref{sec:results}.

\subsection{Harmonic Davidson Algorithm for high-lying roots} \label{sec:harm}

An alternative approach for finding high-lying roots 
is to employ a modified variant of the DA\cite{Davidson1975}.
The so-called harmonic DA\cite{Morgan1991,Morgan1998} is able to find a desired number of target roots
in the vicinity of a user-given excitation energy shift $s$.
In contrast to the CVS ansatz, no additional approximation is introduced here.

\subsubsection{MC-TDA}

The harmonic Davidson (and interior Jacobi-Davidson) algorithm for
solving the generalized eigenvalue problem of symmetric matrices
has been described previously, primarily in the field of solid-state electronic structure theory.\cite{Jordan2012,Tackett2002}
All relevant points to find interior eigenvalues of the MC-TDA
eigenvalue problem are repeated in the following.

To find all interior eigenvalues that are close to shift $s$,
we start with the shifted eigenvalue problem\cite{Tackett2002}
\begin{align}
 ( \bA - s \bSig) \bX =  (\w - s) \bSig \bX \\
 \tilde{\bA} \bX =  \tilde{\w} \bSig \bX \text{.} \label{eq:tda-shift}
\end{align}
In the iterative Davidson-type algorithms, the eigenvectors
\begin{align} \label{eq:evec}
\bX &= \bP\, \bu
\end{align}
 are expanded in set of $M$ trial vectors $\bP = [ \bb_1, \bb_2, \ldots, \bb_M ]$.
This so-called reduced space is expanded in every iteration such that
the approximate eigenvectors in Eq.\ \ref{eq:evec} converge quickly towards the exact solution.

For the harmonic DA, the shifted eigenvalue problem, Eq.\ \ref{eq:tda-shift},
is projected by a different set of trail vectors $\bQ$ from the left
\begin{align}
 \bQ^{\dag} \, \tilde{\bA} \, \bP\, \bu &= \tilde{\w} \, \bQ^{\dag} \, \bSig\, \bP \, \bu \label{eq:tda-proj}\\
 \bQ &= \tilde{\bA} \, \bP \text{.}
\end{align}
Eq.\ \ref{eq:tda-proj} can be expressed entirely in terms of the nonorthogonal $\bQ$ vectors, 
which leads to a shifted and inverted (S\&I) formulation of the generalized eigenvalue problem,
\begin{align}
 \bQ^{\dag} \, \bSig\, \tilde{\bA}^{-1} \, \bQ \, \bu = \frac{1}{\tilde{\w}} \, \bQ^{\dag} \, \bQ \, \bu \label{eq:hd-red-0} \\
 \bQ^{\dag} \, \bSig\, \bP \, \bu = 
\alpha \, \bQ^{\dag} \, \bQ \, \bu  \label{eq:hd-red-1}
\text{.}
\end{align}
By inspecting Eq.\ \ref{eq:hd-red-1}, it is clear that no costly inversion or any solution of a linear system of equations
is ever required.

For reasons that will become apparent later, the S\&I MC-TDA eigenvalue equations in the reduced space, Eq.\ \ref{eq:hd-red-1},
are formulated in terms of linearly transformed Hessian and generalized metric vectors $\bsig$ and $\btau$
\begin{align}
  {\bH}^{\TDA}\, \bu_i &= \alpha_i \, {\bS}^{\TDA}\, \bu_i  \label{eq:hd-red-2} \\
 {H}_{ij}^{\TDA} &= \bsig_i^{\dag} \btau_j -s \, \btau^{\dag}_i \btau_j \\
 {S}_{ij}^{\TDA} &= \bsig^{\dag}_i \bsig_j  -s \, (\btau^{\dag}_i \bsig_j + \bsig^{\dag}_i \btau_j )-s^2 \btau^{\dag}_i\btau_j
\end{align}

To find all those roots that are close to $s$, 
only those eigenvectors $\bu_i$ are considered
that have the largest absolute value $|\alpha_i|$ amongst all $M$ current reduced-space eigenvalues.
As illustrated in Fig.\ \ref{fig3}, the original eigenvalues that
were located formerly in the interior of the spectrum (close to $s$) become
exterior eigenvalues of the S\&I eigenvalue equations and can
be determined by Davidson-type algorithms.

We should note that the reduced-space matrix ${\bH}^{\TDA}$ is non-symmetric
and, thus, complex eigenvalues and eigenvectors may occur in course of the DA.
Nevertheless, the eigenvalues of the untransformed MC-TDA eigenvalue equations
must be real.
If complex roots occur, we follow the approach of Ref.\ \citenum{Jordan2012}
and simply take their absolute values.
Upon convergence, complex roots cannot occur anymore, at least amongst
the set of request target roots.

Another difference between the harmonic and the standard DA
for the MC-TDA generalized eigenvalue equations is
that the Rayleigh quotient
\begin{align}
 \rho_i &= \frac{{\bx}_i^{\dag}\, \bA\, {\bx}_i}{{\bx}_i^{\dag}\, \bSig\, {\bx}_i}
\end{align}
 is not equal to the (harmonic) Ritz values $\theta_i = s + 1/\alpha_i$
before convergence is reached.
This is because the eigenvectors $\bX$ are expanded
in the set of $\bP$ vectors rather than $\bQ$ vectors.
However, the latter are used for the S\&I MC-TDA
eigenvalue problem in the reduced space, Eq.\ \ref{eq:hd-red-0}.\cite{Morgan1998}
Thus, in our implementation $\rho_i$ is computed as
\begin{align}
 \rho_i &=  \frac{{\bx}_i^{\dag}\, {\bsig}_i}{{\bx}_i^{\dag}\, {\btau}_i}
\end{align}
after expanding the (linearly transformed) trial vectors
$\bb$, $\bsig$, and $\btau$ into the full space.
Accordingly, the residual vector of every target root is computed
from $\rho_i$
\begin{align}
 {\br}_i = {\bsig}_i - \rho_i\,{\btau}_i \label{eq:resi}
\end{align}
Trial vectors for the next iteration of the harmonic DA are selected 
and preconditioned as for the standard DA briefly described in Sec.\ \ref{sec:DA}
or elsewhere.\cite{Helmich-Paris2019}

\subsubsection{MC-RPA}

Adapting the harmonic DA for RPA-type generalized eigenvalue problems
is less straightforward because of the diametrical symmetry properties of the $\bsig$ and $\btau$ vectors
that are needed for left projection with $\bQ$.
As shown in Eqs.\ \ref{eq:e-trafo} and \ref{eq:s-trafo}, $\bsig$ vectors preserve the Hermicity $h$ of 
the corresponding trial vectors $b$ while $\btau$ invert $h$.

Let us start with the shifted (MC-)RPA eigenvalue equations for which
the eigenvectors $\bX$ are expanded in a set of trial vectors
\begin{align}
\begin{pmatrix}
 \bsig_{(+)} & - s\,\btau_{(+)} \\
 - s\,\btau_{(-)} &  \bsig_{(-)}
\end{pmatrix}
\begin{pmatrix}
\bu_{(+)} \\
\bu_{(-)}
\end{pmatrix}
=
(\w - s)
\begin{pmatrix}
 0                & - s\,\btau_{(+)} \\
 - s\,\btau_{(-)} &  0
\end{pmatrix}
\begin{pmatrix}
\bu_{(+)} \\
\bu_{(-)}
\end{pmatrix}
\label{eq:rpa-shift}
\end{align}
The matrix at the left-hand side of Eq.\ \ref{eq:rpa-shift} can be readily identified
as the left projector 
\begin{align}
\bQ = 
\begin{pmatrix}
 \bsig_{(+)} & - s\,\btau_{(+)} \\
 - s\,\btau_{(-)} &  \bsig_{(-)}
\end{pmatrix}
\label{eq:q-mat-rpa}
\end{align}
which features a 2-by-2 block structure determined by Hermicity.
Projecting the shifted MC-RPA equation, Eq.\ \ref{eq:rpa-shift}, from the left with $\bQ$
leads to the following
generalized RPA-type eigenvalue problem in the reduced space 
\begin{align} \label{eq:mcrpa-harm-eig-red}
{\bH}^{\RPA} \,
\begin{pmatrix}
\bu_{(+)} \\
\bu_{(-)}
\end{pmatrix}
&=
\alpha_i \, {\bS}^{\RPA} \,
\begin{pmatrix}
\bu_{(+)} \\
\bu_{(-)}
\end{pmatrix}
\\[0.5em]
H_{ij}^{\RPA} &=
\begin{pmatrix}
 - s\,\btau^{\dag}_{(-),i} \btau_{(-),j}  & \bsig_{(+),i}^{\dag}\, \btau_{(+),j} \\
         \bsig_{(-),i}^{\dag}\, \btau_{(-),j} & - s\,\btau_{(+),i}^{\dag} \btau_{(+),j}  &  \\
\end{pmatrix}
\\[0.5em]
S_{ij}^{\RPA} &=
\begin{pmatrix}
 \bsig^{\dag}_{(+),i} \bsig_{(+),j} + s^2\,\btau_{(-),i}^{\dag}\btau_{(-),j} & -s \left( \bsig_{(+),i}^{\dag} \btau_{(+),j} + \btau_{(-),i}^{\dag} \bsig_{(-),j} \right)  \\
  -s \left( \btau_{(+),i}^{\dag} \bsig_{(+),j} + \bsig_{(-),i}^{\dag} \btau_{(-),j} \right) 
& \bsig^{\dag}_{(-),i} \bsig_{(-),j}+ s^2 \btau^{\dag}_{(+),i} \btau_{(+),j}
\end{pmatrix}
\end{align}
that is solved in every iteration of in the harmonic DA.

In contrast to the generalized RPA eigenvalue problem that
is solved with the standard DA, Eq.\ \ref{eq:mcrpa-eig-red},
non of the Hermicity blocks vanishes in the eigenvalue equations \ref{eq:mcrpa-harm-eig-red}.
Thus, the computational costs for solving Eq.\ \ref{eq:mcrpa-harm-eig-red} are much higher
than those of Eq.\ \ref{eq:mcrpa-eig-red}.
However, for most applications only it is sufficient to find only a few roots
and, thus, the costs for calculating the MC-RPA sigma vectors are usually
much higher than those for solving the eigenvalue equations in the reduced space.

The reduced-space matrix ${\bH}^{\RPA}$ is non-symmetric as in the
MC-TDA case and, again, complex eigenvalue may occur.

\section{Computational Details} \label{sec:compdet}

All calculations were performed with a development version of ORCA
- an \textit{ab initio}, DFT and semi-empirical SCF-MO package.\cite{Neese2012,Neese2018}
The CVS approximation and the harmonic DA have been implemented 
in the \texttt{orca\_mcrpa} module.\cite{Helmich-Paris2019}
More technical details on \texttt{orca\_mcrpa} can be found in Ref.\ \citenum{Helmich-Paris2019}.
The energy shift used for the S\&I approach was derived from other calculations
or experimental values.
Please note that the exact value of the shift is not relevant.
Taking a slightly different shift will at most lead to shifting the energy window range 
of the converged roots.
If many roots are requested, as it is usually the case when simulating
UV/Vis and XAS spectra, a slight shift in the energy window should
be of no concern.

All core orbitals were considered in the calculations whenever the CVS approximation
was not used.
To account for scalar-relativistic effects, we employed the spin-free second-order 
Douglas--Kroll--Hess (DKH2) Hamiltonian.\cite{Douglas1974,Jansen1989}
For all DFT calculations we used the B3LYP exchange-correlation functional\cite{Becke1988,Lee1988,Becke1993}
and the unrestricted Kohn-Sham ansatz.
In the subsequent TD-DFT calculation the CVS approximation was applied,
i.e.\ the orbital window for the occupied orbitals was restricted
to a specific set of core orbitals.

%
%
The RI approximation has been used for all MR methods.
The auxiliary basis set was automatically generated by the AutoAux
program available in ORCA.\cite{Stoychev2017}

Note that we have used the electric dipole approximation when computing
oscillator strengths for all XA spectra.

For reasons of consistency with the study of Chantzis \textit{et al.},\cite{Chantzis2018} 
the uncontracted cc-pVTZ basis set\cite{Balabanov2005} was used for all calculations on the Manganese atom.
For the CASSCF ground state energy and MC-RPA
calculations on Manganese, a CAS with
 five electrons in five (d) orbitals in a sextet state was chosen.
The CAS-CI and NEVPT2\cite{Angeli2001,Angeli2001b,Angeli2002} results were obtained as described in Ref.\ \citenum{Chantzis2018},
i.e.\ the 1s, 2p, or 3p core orbitals, respectively, of the CAS-(5,5)-SCF calculation were rotated
right below the active 3d orbitals;
then the active space was augmented by the core electrons and orbitals (CAS(7,6) for K-edge, CAS(11,8) for L- and M-edge)
in the succeeding CAS-CI and NEVPT2 calculation.
The strongly-contracted NEVPT2 variant was employed in this study.
While the CVS CAS-CI calculation reuses the CAS-(5,5)-SCF without any modification,
in NEVPT2 the MOs get canonicalized for each state by diagonalizing the state-specific total Fock matrix $\mathbf{F}$
that depends on the CAS-CI solution vector in its active part.\cite{Roos1980}

For all calculations on the permanganate ion $\text{MnO}_4^-$,
a perfect tetrahedral structure was assumed with an experimental bond length of 1.629~\AA.\cite{Palenik1967}
A customized version of the def2-QZVPP basis set\cite{Weigend2005} was used
with special contraction coefficients for the DKH2 Hamiltonian.\cite{PantazisDKH}
An active space with all Mn 3d and all O 2p electrons and orbitals, CAS (24,17),
was chosen for all the CASSCF and MC-RPA calculations.
A different active space, CAS(8,9), was employed for the CVS CAS-CI and NEVPT2 calculations
that contains all O 1s and the Mn 3d electrons and orbitals.
Those calculations used the orbitals from the preceding CAS-(24,17)-SCF calculation.
We have again used state-specific canonical orbitals as for the Manganese calculations.

\section{Results and discussion} \label{sec:results}

\subsection{Accuracy of the CVS approximation} \label{sec:mn-cvs}

The accuracy of the CVS approximation is investigated for
the K-, L-, and M-edge transition of the Manganese atom.
Those transition are dominated by a one-electron promotion
from the 1s (K), 2p (L), and 3p (M) core shell into the half-filled
valence 3d shell.
The MC-RPA excitation energies for those transitions are compiled in Tab.\ \ref{tbl1}
along with the TD-DFT B3LYP, CAS-CI, and NEVPT2 results.
Excitation energies are provided as differences to the NEVPT2 results.
Note that only for the K-edge an experimental reference is provided.
A realistic simulation and assignment of L- and M-edge spectra\cite{Arp1992}
 requires inclusion of spin-orbit coupling due to the pronounced splitting of core p shells, 
which is currently not available, at least, for the CASSCF LR implementation.

As can be seen from Tab.\ \ref{tbl1}, the transition energies from 
the CASSCF type methods, i.e.\ CAS-CI, CVS MC-RPA and S\&I MC-RPA, are
systematically blue shifted,
while those from CVS TD-DFT are systematically too small.
This trend is fully in line with benchmark results on valence excitations energies\cite{Helmich-Paris2019a}
and can be readily attributed to an insufficient treatment of
dynamic electron correlation.

Much more interesting is the accuracy of the CVS approximation in comparison
to the exact S\&I approach for high-lying MC-RPA excitation energies.
In case of K-edge transitions, which involve single-electron excitations from 
the lowest 1s orbital, the difference between CVS  and S\&I is roughly $10^{-4}$~eV
and many orders of magnitude below the inherent error of the electronic structure method
or the basis set incompleteness error.

A completely different observation can be made for the M-edge spectra
for which electrons in the 3p shell get promoted into the valence 3d shell.
These excitations occur in the presence of electromagnetic radiation from
UV part of the light spectrum rather than the (soft) X-ray part.
For those transitions, we observe a complete breakdown of the
CVS approximation.
CVS and S\&I MC-RPA excitation energies differ by up to 3.5~eV, which is
twice as large as the energetic difference to the NEVPT2 results.
Though exhaustive benchmarking was not made in the current study,
we may conclude that CVS is an excellent approximation 
for K-edge transitions.
In case of transitions induced by UV or low-energy X-ray radiation
the S\&I approach based on the harmonic DA has to be
used to prevent severe errors.

\subsection{O K-edge XAS of $\text{MnO}^-_4$}

To demonstrate the feasibility of MC-RPA for simulating XAS spectra of molecular systems,
the oxygen pre- and near K-edge XA was simulated by computing 50 roots 
using either the CVS approximation or the S\&I approach with a shift of 530~eV.
With both Davidson-type implementations (standard and harmonic DA)
the Frobenius norms $||\br||$ of the residual vectors, Eq.\ \ref{eq:resi}, converged smoothly below
a threshold of $10^{-4}$ within 8 and 11 iterations, respectively,
as shown in Fig.\ \ref{fig4}.
Interestingly, the S\&I algorithm converges faster though the search space for the
eigenvectors is much larger.
Note that the curve flattening of the minimum and average $||\br||$ near convergence 
is not a sign of convergence issues, but rather
caused by an increasing number of converged eigenpairs that are then 
locked by the DA and not improved further.

Concerning the accuracy of the two approaches, the errors introduced 
by the CVS approximation are well below 0.1~eV for all roots and irrelevant for any practical purpose.
Note that a much more significant deviation between the CVS approximation
and the exact S\&I approach will most likely be observed when simulating L- and M-edge
spectra as seen in Sec.\ \ref{sec:mn-cvs}.
The computational savings due to the CVS approximation are huge for such
a large active space CAS(24,17) MC-RPA calculation that expands the
configuration part of the eigenvectors in more than $8.8 \times 10^6$~CSFs.
In case of CVS MC-RPA, only the CI problem of the ground must be solved and
the calculation completed after 30~min using 20~MPI processes on an
Intel Haswell node (Intel{\textregistered} Xeon{\textregistered} CPU E5-2687W v3 @ 3.10~GHz).
Conversely, it took roughly 17~h using 24~MPI processes on an
Intel Broadwell node (Intel{\textregistered} Xeon{\textregistered} CPU E5-2687W v4 @ 3.00~GHz)
until the S\&I MC-RPA calculation finished. 
For the  S\&I MC-RPA calculation each vector is expanded in the complete space of orbital rotations
and CSFs.
When considering both accuracy and efficiency, the CVS approximation should
be clearly preferred to the S\&I approach for CASSCF LR K-edge spectra.

The O K-edge XAS MC-RPA spectrum (S\&I) of $\text{MnO}_4^-$ is shown in the
top panel of Fig.\ \ref{fig5}.
The two characteristic pre-edge peaks correspond to a transition from the O 1s
orbitals into the empty $\text{t}_2^*$ and $\text{e}^*$ orbitals.
Peaks with higher energy in the MC-RPA spectrum belong to the near-edge fine structure region
that have so far not been reported by experimental studies.
In comparison to the known experimental values of the K pre-edge region,\cite{Minasian2013}
the two MC-RPA peaks are blueshifted by roughly 10.5~eV.
The splitting of the two peaks of 0.85~eV is too small compared to the experimental value of 1.50~eV.
Furthermore, our assignment of the K pre-edge peaks in the MC-RPA spectrum
contradicts the assignment made in the experimental study of Ref.\ \citenum{Minasian2013}.
In our MC-RPA calculation, transitions into the $t_2^*$ orbitals are energetically below
the transitions into the $e^*$ orbitals, which seems non-physical for a 
tetrahedrally coordinated TM complex. 
Still, a physically sound orbital ordering is revealed
when inspecting the CAS-(24,17)-SCF orbital coefficients and pseudo orbital energies.
Thus, we suspect a wrong ordering of the MC-RPA pre-edge transitions
caused by the lack of dynamical electron correlation rather than
a wrong interpretation made in the experimental study of Ref.\ \citenum{Minasian2013}.
The presumption is supported when considering the CAS-CI and NEVPT2 K-edge XA spectrum
presented in the middle and lower panel of Fig.\ \ref{fig5}, respectively.
While the $\text{O 1s} \to \text{t}_2^*$ transition is slightly below the one of 
$\text{O 1s} \to \text{e}^*$ for the CAS-(8,9)-CI calculation, the physically sound
ordering in accordance with the ligand-field picture of tetrahedrally coordinated
TM complexes is observed in the NEVPT2 calculation.
Though the ordering seems to be correct in the NEVPT XA spectrum,
the transition energies are significantly redshifted and feature a much larger absolute
deviation from the experimental peaks than with MC-RPA.
Furthermore, those state-specific NEVPT2 calculations show a significant splitting
of the  $\text{O 1s} \to \text{e}^*$ and $\text{O 1s} \to \text{t}_2^*$,
which seems to be unphysical and not in line with the experimental spectrum of Ref.\ \citenum{Minasian2013}.
Nevertheless, our MC-RPA XA spectrum gave the least deviation from the experimental spectrum
except of the questionable ordering of the two pre-edge peaks.
We have faith that future computational studies with more accurate MR electronic structure methods
will improve the agreement between theory and experiment.

\section{Conclusions}

In the present work, we have introduced two approaches for simulating
XA spectra with the CASSCF LR method also known as MC-RPA.
The first approach employs the well-known CVS approximation, which
is predominantly used by other electronic structure methods for simulating
X-ray spectra.
The second ansatz (S\&I) uses the harmonic DA for interior eigenvalues
that lie close to a user-given shift. Similar algorithms for interior eigenvalues were used already 
in the context of computational XAS for SR methods as TD-DFT\cite{Kasper2018} and
EOM-CC\cite{Zuev2015}, but not yet for MR LR approaches.

Both implementations of the corresponding Davidson-type algorithm 
show fast convergence. Even faster convergence is observed for the harmonic DA
S\&I MC-RPA method for the cases considered in this work.

Furthermore, we could show in a proof-of-principle application to the Manganese atom that the
additional errors for excitations energies introduced by the CVS approximation compared to the exact S\&I approach
are negligible for K-edges, but become larger than the methodological error
of the MC-RPA method when computing M-edges.
Hence, whenever simulating M-edge spectra, the S\&I approach becomes unavoidable.

Nevertheless, in case of large active-space K-edge calculations, the computational savings of
the CVS approximation are huge as the configuration part of the response vectors
is completely neglected.
For such calculations, the CVS approximations should be employed.

The O K-edge XA spectrum of $\text{MnO}_4^-$ was simulated with three MR methods, i.e.\ 
our one-step MC-RPA approach and the multi-step CAS-CI / NEVPT2 methods.
For all those three MR methods, MC-RPA showed the best agreement with the experimental spectrum.
However, the order of the two computed pre-edge peaks seems to be questionable.
We suspect that MR LR methods that can account for dynamic correlation\cite{Sokolov2018,Koehn2019}
could provide an even better agreement with the experimental spectrum.

Future developments will focus mainly on the following aspects:
(i) inclusion of dynamic correlation to obtain higher accuracy, 
(ii) treatment of degenerate
open-shell ground states that suffer from symmetry breaking,
(iii) and  the inclusion of spin-orbit coupling to simulate L- and M-edge
spectra.
Currently, we are working actively on the first two aspects in our laboratory.

\section{Acknowledgments}

The author thanks cordially Frank Neese, Dimitrios Manganas, and Mihail Atanasov for fruitful discussions.

\section{Funding Information}

Financial support by the Max Planck society is gratefully acknowledged.

\clearpage


\bibliography{references}   


\clearpage

\begin{figure}[h]
\caption{\label{fig1} Schematic block structure of the full MC-TDA Hessian.
 Orbital rotations between the inactive, active, and virtual MO spaces are labeled V-O, A-O, and V-A.
 CI denotes the configuration part.}
\end{figure}

\begin{figure}[h]
\caption{\label{fig2} Schematic block structure of the CVS MC-TDA Hessian. For further details see Fig.\ \ref{fig1} and text.}
\end{figure}

\begin{figure}[h]
\caption{\label{fig3} Relation between the eigenvalues of the original and the S\&I 
 eigenvalue problem. Eigenvalues close to shift $s$ are relocated from the interior of
 the original eigenvalue spectrum to the exterior of the
 S\&I eigenvalue spectrum.}
\end{figure}

\begin{figure}[h]
\caption{\label{fig4} Convergence of the minimum, maximum, and average
residual Frobenius norm $||\br||$ of a 50 root CVS and S\&I MC-RPA calculation.
Convergence threshold for $||\br||$ was $10^{-4}$.
See text for further details.
}
\end{figure}

\begin{figure}[h]
\caption{\label{fig5} O K-edge XA spectrum of $\text{MnO}_4^{-}$ with
S\&I MC-RPA, CAS-CI, and NEVPT2. Dashed lines are experimental peak maxima of
the $1s \to e^*$ and $1s \to t_2^*$ pre-edge transitions.
A Gaussian broadening of 0.5~eV is employed which corresponds to
to the spectral resolution of the experimental spectrum given in Ref.\ \citenum{Minasian2013}.}
\end{figure}



\clearpage

\vspace*{0.1in}   
\begin{center}
\includegraphics[width=0.9\columnwidth,keepaspectratio=true]{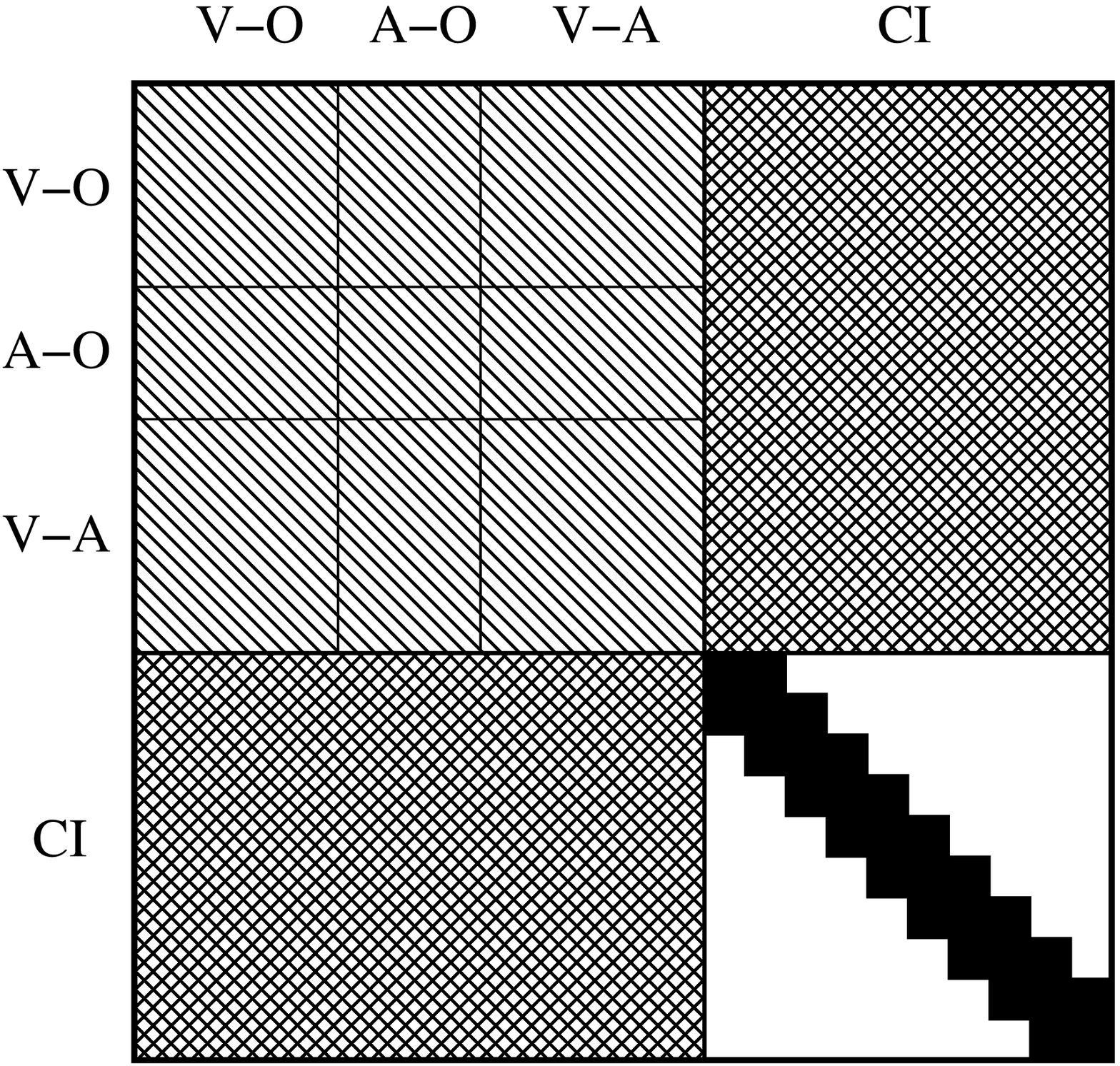}
\end{center}
\vspace{0.25in}
\hspace*{3in}
{\Large
\begin{minipage}[t]{3in}
\baselineskip = .5\baselineskip
Figure 1 \\
B.\ Helmich-Paris  \\
Int. J.\ Quant.\ Chem.
\end{minipage}
}

\vspace*{0.1in}   
\begin{center}
\includegraphics[width=0.9\columnwidth,keepaspectratio=true]{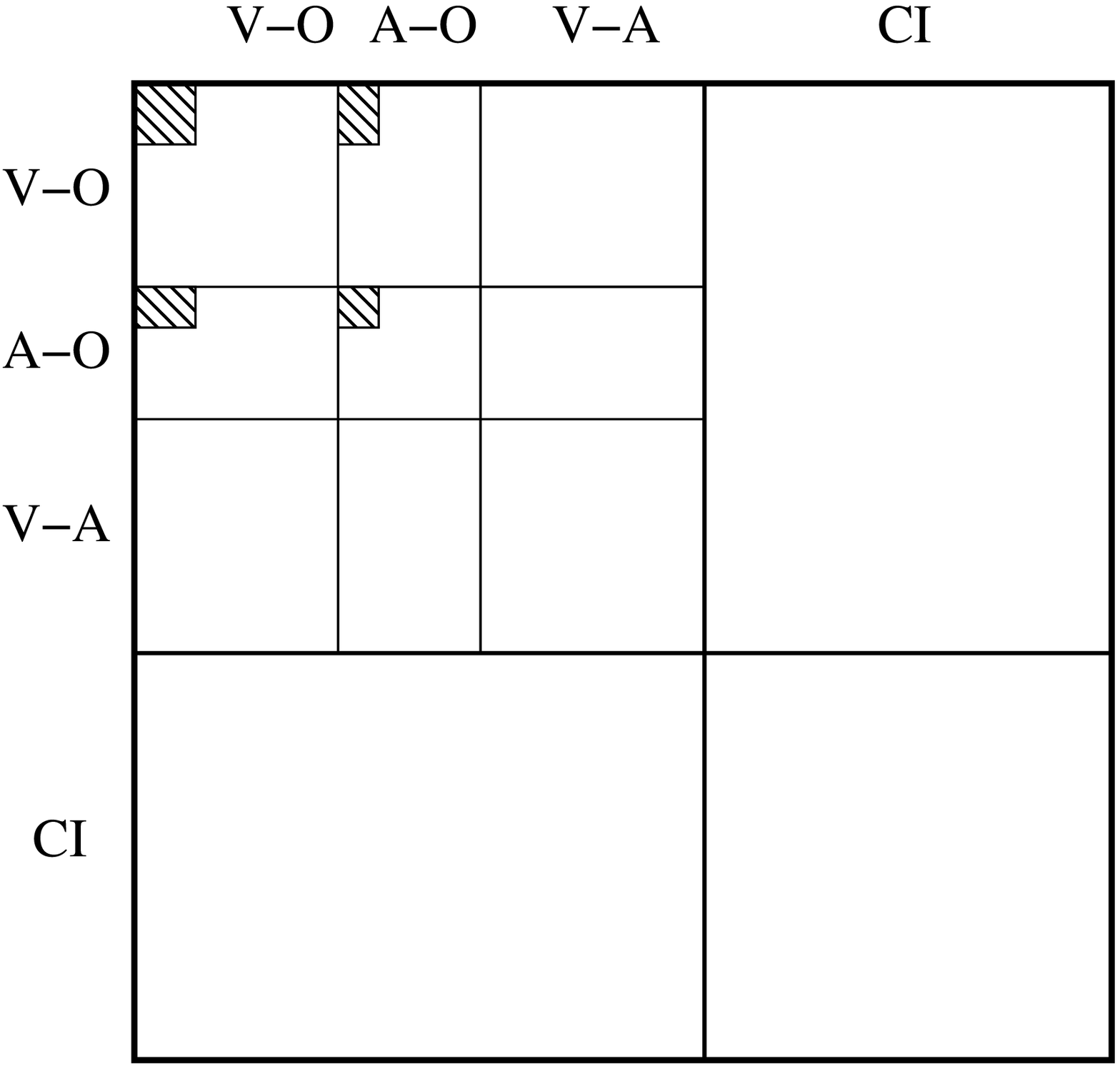}
\end{center}
\vspace{0.25in}
\hspace*{3in}
{\Large
\begin{minipage}[t]{3in}
\baselineskip = .5\baselineskip
Figure 2 \\
B.\ Helmich-Paris  \\
Int. J.\ Quant.\ Chem.
\end{minipage}
}

\vspace*{0.1in}   
\begin{center}
\includegraphics[width=0.9\columnwidth,keepaspectratio=true]{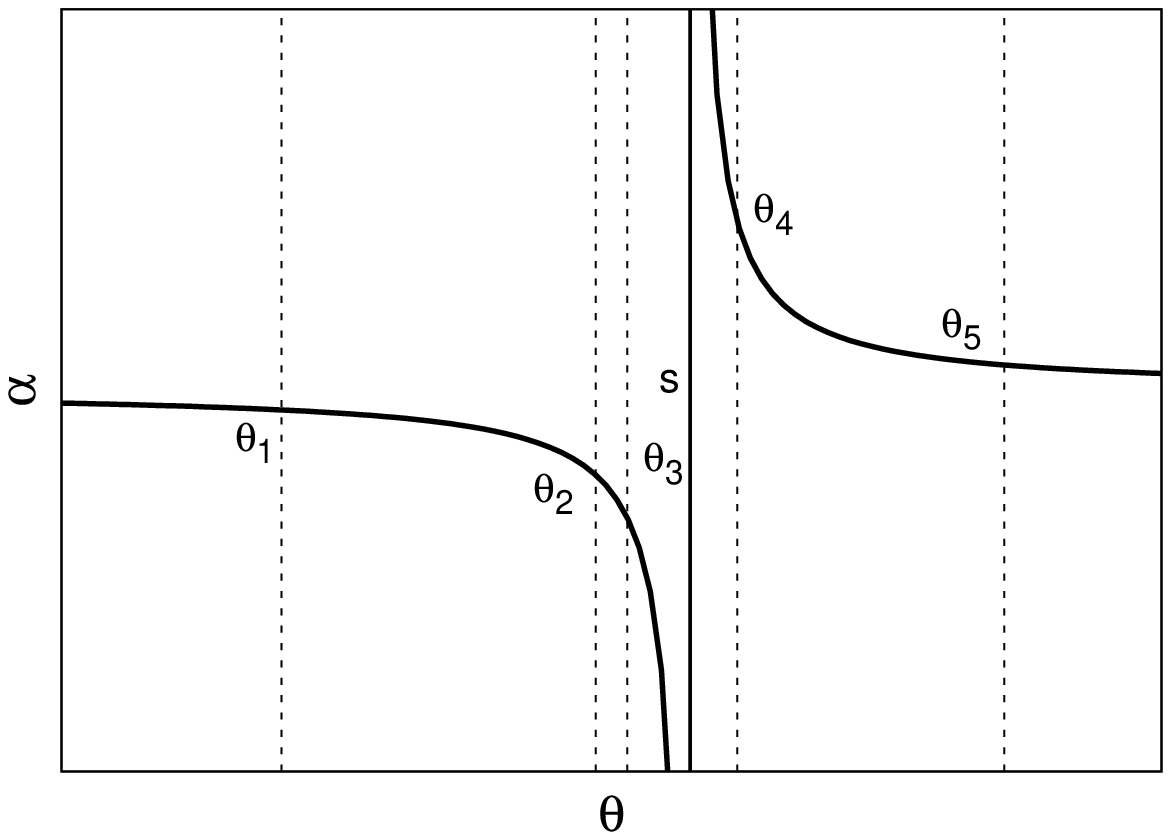}
\end{center}
\vspace{0.25in}
\hspace*{3in}
{\Large
\begin{minipage}[t]{3in}
\baselineskip = .5\baselineskip
Figure 3 \\
B.\ Helmich-Paris  \\
Int. J.\ Quant.\ Chem.
\end{minipage}
}

\vspace*{0.1in}   
\begin{center}
\includegraphics[width=0.9\columnwidth,keepaspectratio=true]{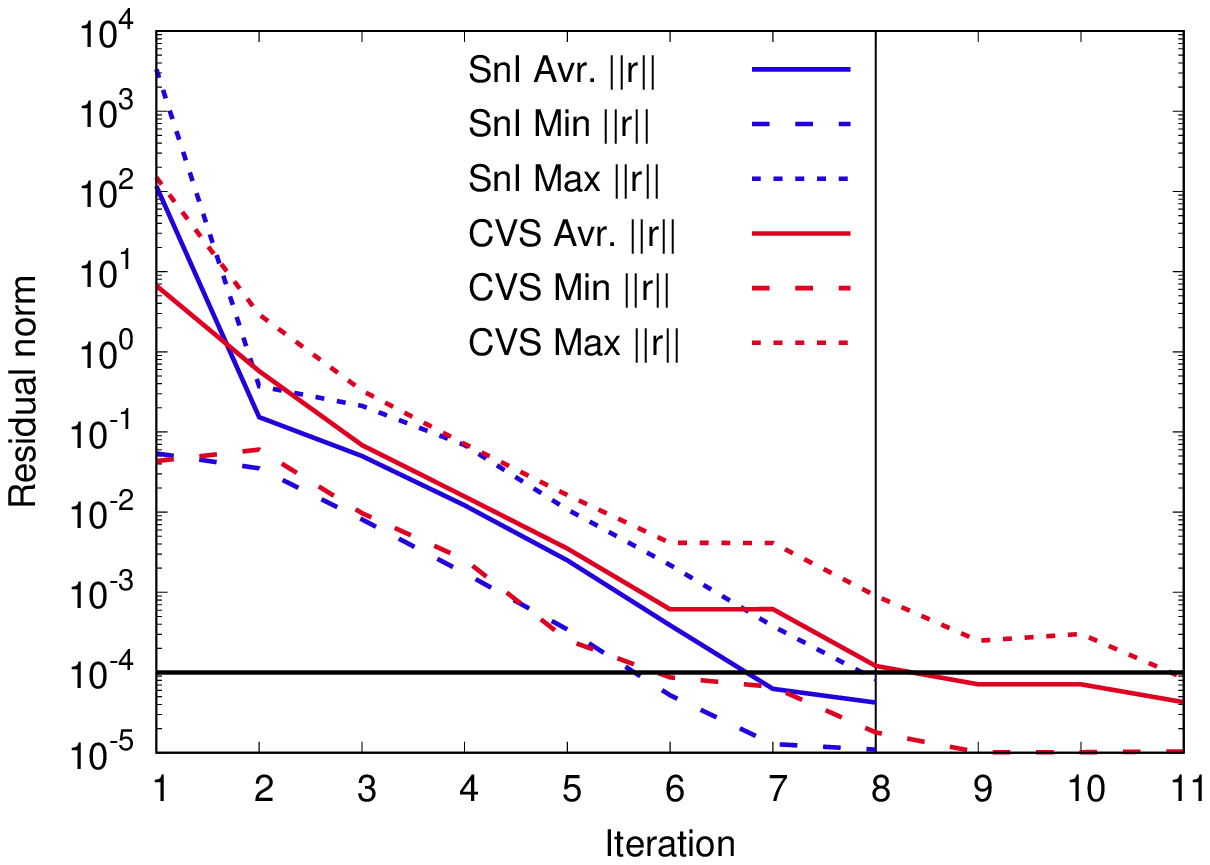}
\end{center}
\vspace{0.25in}
\hspace*{3in}
{\Large
\begin{minipage}[t]{3in}
\baselineskip = .5\baselineskip
Figure 4 \\
B.\ Helmich-Paris  \\
Int. J.\ Quant.\ Chem.
\end{minipage}
}

\vspace*{0.1in}   
\begin{center}
\includegraphics[width=0.9\columnwidth,keepaspectratio=true]{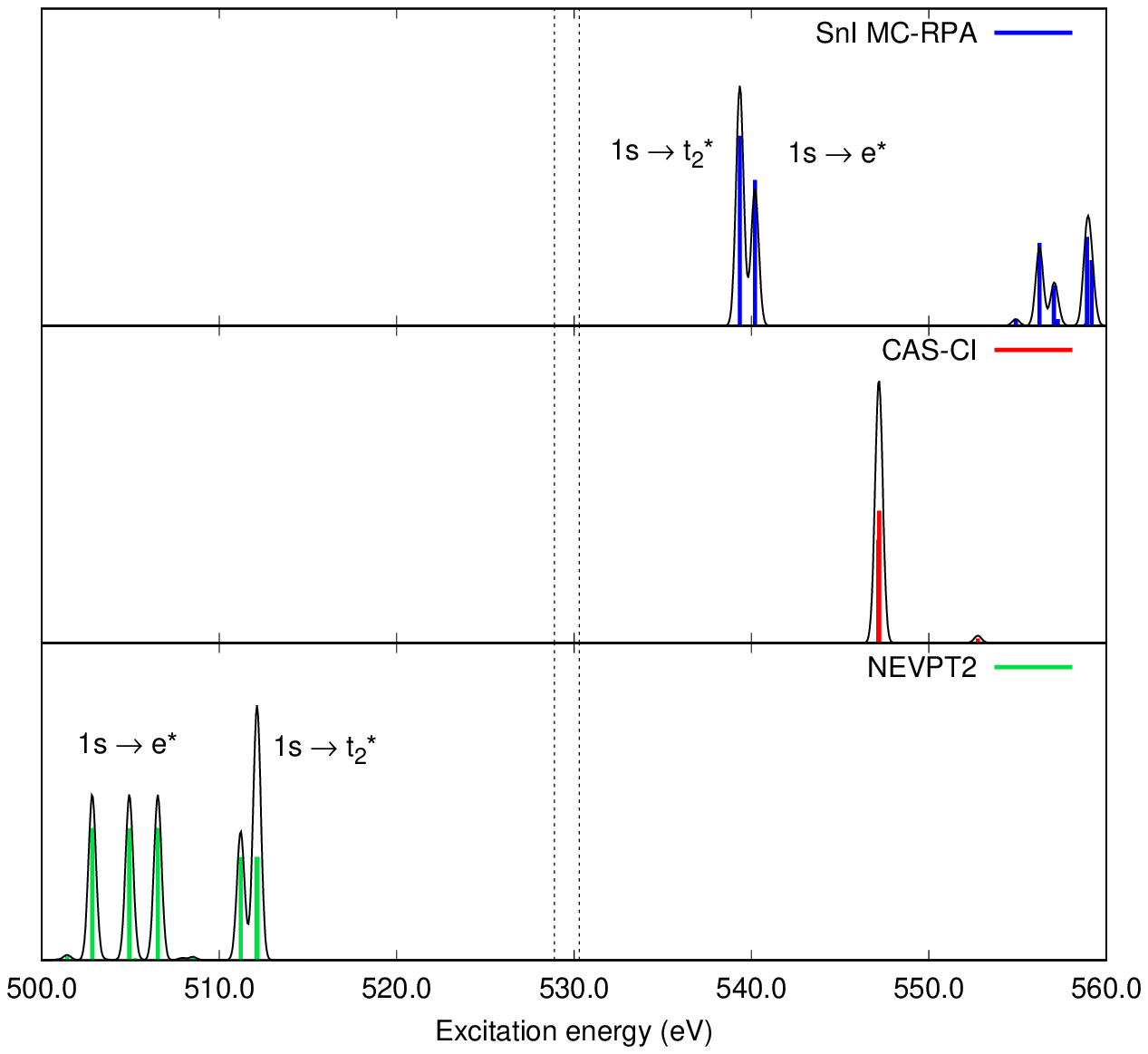}
\end{center}
\vspace{0.25in}
\hspace*{3in}
{\Large
\begin{minipage}[t]{3in}
\baselineskip = .5\baselineskip
Figure 5 \\
B.\ Helmich-Paris  \\
Int. J.\ Quant.\ Chem.
\end{minipage}
}

\clearpage
\begin{table}
\begin{tabular}{crrrrrrr}\hline
\textbf{State} & \textbf{Expt.} & \multicolumn{5}{c}{\textbf{Methods}} \\
     &   & \textbf{NEVPT2} & \textbf{$\Delta_{\text{CAS-CI}}$} & \textbf{$\Delta_{\text{B3LYP}}$} & \textbf{$\Delta_{\text{CVS MC-RPA}}$} & \textbf{$\Delta_{\text{S\&I MC-RPA}}$} & \textbf{CVS - S\&I} \\ \hline
\multicolumn{8}{c}{K-edge} \\
   ${^6}D$ & 6545.88 & 6535.70 & +40.67 & -84.71 & +39.82 & +39.82 & +1.2~$10^{-4}$ \\[1.0em]
\multicolumn{8}{c}{L-edge} \\
   ${^6}F$ &  /      & 640.87 & +13.24 & -11.02 & +12.70 & +12.67 & +3.0~$10^{-2}$ \\
   ${^6}D$ &  /      & 641.76 & +13.35 & -10.40 & +12.86 & +12.85 & +4.5~$10^{-3}$ \\
   ${^6}P$ &  /      & 642.96 & +12.53 &  -7.61 & +12.39 & +12.40 & -4.3~$10^{-3}$ \\[1.0em]
\multicolumn{8}{c}{M-edge} \\
   ${^6}F$ &  /      & 47.59 & +0.58 & -0.87 & -0.16 & -0.31 & +1.5~$10^{-1}$ \\
   ${^6}D$ &  /      & 48.64 & +1.02 & -1.21 & +0.31 & +0.30 & +6.9~$10^{-3}$ \\
   ${^6}P$ &  /      & 50.61 & +3.66 & +1.38 & +1.99 & -1.50 & +3.5~$10^{ 0}$
\end{tabular}
\caption{\label{tbl1} K-, L-, and M-edge XA transitions of the Manganese atom with different methods
as well as the experimental K-edge are provided in eV.
Differences $\Delta$ are given with respect to the NEVPT2 results.
For further details see text.
}
\end{table}

\end{document}